\documentclass[aps,prd,preprint,superscriptaddress,showpacs,nofootinbib]{revtex4-1}
\usepackage{graphics}
\usepackage{epsfig}
\usepackage{latexsym}
\usepackage{colordvi}
\usepackage{amsmath}
\usepackage{amssymb}

\begin{document}

\preprint{\parbox[b]{1in}{ \hbox{\tt PNUTP-20/A05}  }}

\title{Parity violation and new physics in superconductors}

\author{Deog Ki Hong}
\email[E-mail: ]{dkhong@pusan.ac.kr}
\affiliation{Department of Physics, Pusan National University,
             Busan 46241, Korea}

\vspace{0.1in}

\date{\today}

\begin{abstract}
We propose a new method, using the Andreev reflection at superconductors, to measure parity violation induced by the standard electroweak theory, which in turn constrains the possible parity-violating effects of new physics. 
The weak neutral currents induce  parity-violating, marginal effective  operators, though quite tiny, in superconductors. We estimate their effects on superconducting gaps and propose a method to measure the parity-violation  from the spin polarization effect,  when electrons or holes get Andreev-reflected at the interface between normal metal and a superconductor. 
Such polarization effects might be comparable to the atomic parity violation and thus naturally give an interesting bound on certain models of new physics, that couples to electrons, such as Majorana mass of active neutrinos or doubly charged Higgs. 
\end{abstract}


\maketitle

\newpage

\section{Introduction}
Weak interaction is the only fundamental force, known to break parity. The parity is broken at the fundamental level, as the standard electroweak theory, that unifies the weak and electromagnetic forces, is chiral. 
The striking phenomena of parity non-conservation in the electroweak process are prominent in deep inelastic scattering of neutrinos off nucleons, that confirms the existence of the weak neutral currents, as predicted by the standard model of particle physics~\cite{paschos,Amaldi:1987fu}. However, the pari	`ty violation effect is very difficult to detect at long distances, because the weak interaction is of very short range. Over the years since a possibility of measuring parity-violation in heavy atoms was noted by Bouchiat and Bouchiat~\cite{Bouchiat}, a set of ingenious experiments has been performed and verified  such parity violation in atoms, by measuring the difference in the transition rates of atoms induced by polarized photons~\cite{barkov,Noecker:1988ys,Wood}. 

The observation of atomic parity violation (APV) has not only confirmed  the standard electroweak theory but also highly constrains possible new physics beyond the standard model~\cite{marciano_rosner,oai:arXiv.org:hep-ph/9705435,iw}, which might produce a large parity-violation at low energy. In this letter we propose another method to measure the parity violation at low energy, using superconductors.  Since superconductors exhibit macroscopically the quantum phenomena of Bose-Einstein condensation of Cooper-pairs, the microscopic violation of parity can be dramatically amplified in superconductors. As the underlying force of Cooper pairing in superconductors preserves the parity,  any parity violating effect on Cooper-pairing can therefore lead to observable consequences. We first estimate the effect of the parity-violating standard electroweak theory in the Cooper-paring gap and then propose as its observable consequences  the spin-polarization of electrons or holes, when they get Andreev-reflected at the interface of metal and superconductor. Because of the parity-violation the superconducting gap is helicity-dependent to result in an order of $10^{-15}$ difference in the gap of two different helicity eigenstates, which might be comparable to APV in the dipole matrix element, $10^{-11} |e|a_0$~\cite{Blundell:1990ji}, if the Bose-Einstein condensation effect is taken into account as in the Andreev reflection.  

Finally we consider the new physics effects due to the Majorana mass of
active neutrinos and the extended Higgs sector of the standard model, whose effects on the superconductivity possibly enhance the parity-violation. 
Majorana neutrino mass will have an observable consequence in a
system where the lepton number is spontaneously broken as in (electronic) superconductors. We also show that new physics models such as the minimal model of  type \uppercase\expandafter{\romannumeral2\relax} seesaw~\cite{Magg:1980ut,Schechter:1980gr}  or models with Majorana particles can be stringently constrained.

\section{Effective theory}

The low energy effective theory of electrons of mass $m_*$ in metal is described by a Lagrangian
density with a chemical potential $\mu$
\begin{equation}
{\cal L}_{\rm eff}= \Psi^{\dagger}\left[iD_t+
\frac{1}{2m_*}\left(\vec {\sigma}\cdot \vec {D}\right)^2+\mu\right]\Psi
+\frac{g}{\Lambda^2}\left(\Psi^{\dagger}\sigma^2\Psi\right)^2+\cdots
\label{eff}
\end{equation}
where $\Psi=(\Psi_{\uparrow},\Psi_{\downarrow})^T$ is the two-component electron field of spin up and down,
$\sigma^i$'s ($i=1,2,3$) are the Pauli matrices,  and $D_{\mu}=\partial_{\mu}+ie\,A_{\mu}$ is
the covariant derivative of quantum electrodynamics. The four-Fermi interaction is the leading term due to phonon
exchange interactions of electrons, expanded in powers of derivatives. The ultraviolet cut-off $\Lambda$ is 
the Debye energy or the characteristic energy of phonons, $\hbar\,\omega_D$, below which the effective theory is valid. 
The relevant symmetry of the system for our discussion is the spin symmetry and the electromagnetism,
${\rm SU}(2)_S\times {\rm U}(1)_{\rm em}$.

Since the attractive four-Fermi interaction is marginally relevant for electrons
with opposite momenta near the Fermi surface, the four-Fermi interaction derives 
the electrons to form Cooper-pairs and condense, 
opening a superconducting gap at the Fermi surface that breaks not only the ${\rm U}(1)_{\rm em}$ but also the electron numbers. 
Since the bulk of electrons except  those lying near the Fermi surface
are irrelevant in describing the low-energy properties of conductors, we may expand
the electron field to describe modes near the Fermi surface as, following the high density effective theory
(HDET)~\cite{Hong:2000tn},
\begin{equation}
\Psi(x)=\sum_{\vec {v}_F}\,e^{i\vec {p}_F\cdot \vec {x}}\,\psi(\vec {v}_F,x),
\end{equation}
where $\vec {p}_F=m_*\vec {v}_F$ is the Fermi momentum and the summation is over the
patches of the Fermi surface, labeled by the Fermi velocity $\vec {v}_F$.

In the non-relativistic formalism the chirality of electrons is nothing but their helicity. To see this we consider a four-component spinor in the basis of chirality,
\begin{equation}
\psi(x)=\begin{pmatrix}\psi_L\\
\psi_R
\end{pmatrix}\,,
\end{equation}
where the left (right)-handed spinors are defined as $\psi_{L}=\frac12(1+\gamma_5)\psi$ and $\psi_R=\frac12(1-\gamma_5)\psi$\, respectively. Each spinor has both negative and positive energy states. In the dense fermionic matter the positive energy states near the Fermi surface are give as 
 \begin{equation}
\psi(\vec{v_F},x)=\frac{1+\vec\alpha\cdot \hat{v}_F}{2}e^{-i\vec{p}_F\cdot\vec x}\,\psi(x)\,,
\end{equation}
where $\vec\alpha=\gamma_0\vec\gamma$ is the spin operator and ${\hat v}_F=\vec v_F/|\vec v_F|$~\cite{Hong:2000tn}. We see that therefore the left  (right)-handed
positive-energy spinor is the electron near the Fermi surface with spin (anti)-parallel to the Fermi momentum. In other words, $\psi_L(\vec v_F,x)=\psi_{\uparrow}(\vec v_F,x)$ and 
$\psi_R(\vec v_F,x)=\psi_{\downarrow}(\vec v_F,x)$\,.

As the Cooper pairs in ordinary superconductors are spinless, the condensate
takes the following form in the basis of chirality:
\begin{eqnarray}
\left<\psi_L^T(-\vec {v}_F,x)\psi_L(\vec {v}_F,x)\right>=K_L\,e^{i\phi_L/F_L},
\quad \left<\psi_R^T(\vec {v}_F,x)\psi_R(-\vec {v}_F,x)\right>=K_R\,e^{i\phi_R/F_L},
\end{eqnarray}
where the left-handed (right-handed) electrons have spin opposite
(parallel) to their Fermi momenta.
In general, the condensates of left-handed electrons need not be equal to those of
right-handed electrons. But,
since the dynamics of electrons in conductors is governed by
the phonon-exchange interaction and
the electromagnetic interaction, which  are parity-invariant,
we have $K_L=K_R\,(\equiv K)$ and $F_L=F_R\,(\equiv F)$.
When the Cooper-pairs condense, a superconducting gap opens
for electrons of both helicities at the Fermi surface ($i=L,R$),
\begin{eqnarray}
E_i(\vec p)=\pm\sqrt{\left(\frac{{\vec p}^2}{2m}-\frac{{\vec p_F}^2}{2m}\right)^2+\Delta_i^2}\,,
\quad\Delta_i=\left|K_i\right|/p_F^2.
\end{eqnarray}
The superconducting gap is  nothing but a Majorana mass of
electrons at the Fermi momentum that does not conserve the electron number~\cite{Hong:2000ng}.
The mass term for superconducting electrons is given in HDET as
\begin{eqnarray}
{\cal L}_{\rm mass}=-\Delta_L\,
\bar\psi^C_{L}(\vec {v}_F,x)\psi_{L}(\vec {v}_F,x)
-\Delta_R\,\bar\psi^C_{R}(\vec {v}_F,x)\psi_{R}(\vec {v}_F,x) + {\rm h.\,c.},
\label{mass}
\end{eqnarray}
where the summation over the patches is suppressed and $\psi^C\equiv C\bar\psi^T$ is the charge conjugate field~\cite{Hong:2000ck}.
Since the condensates preserve the ${\rm SU(2)}_S$ spin symmetry, $K_L=K_R$ or $\Delta_L=\Delta_R\,(\equiv\Delta)$, 
only the vectorial combination of the
small fluctuations of the condensate constitutes the supercurrent\footnote{The difference between two phases will constitute the axion currents.}
\begin{eqnarray}
j_{\mu}(x)=F\,i\partial_{\mu}\phi(x),\quad\phi=\phi_L+\phi_R.
\end{eqnarray}
The superconducting gap makes
the small fluctuations of Cooper pairs be the supercurrent excitations and any
magnetic fields are expelled outside the superconductor, known as the Meissner effect.

\begin{figure}
\includegraphics[scale=0.5]{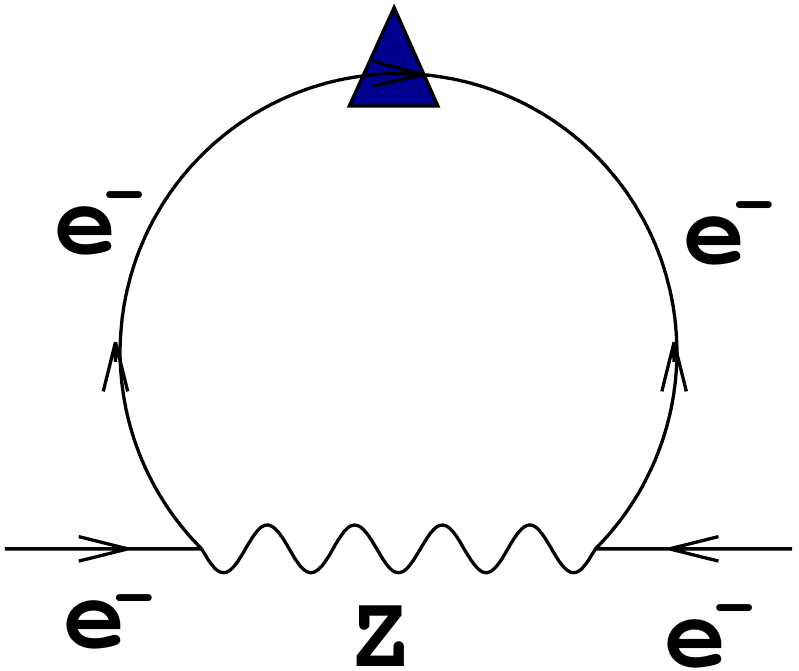}%
\caption{The  triangle denotes
the superconducting gap of electrons and the  wiggly lines denote the weak $Z$ boson.}
 \label{wnc}
\end{figure}

Since in addition to the phonon exchange interaction the electrons do interact with each other by the Z-boson exchange,
the left and right-handed electrons will open gaps slightly differently, signifying the parity-violation. 
The weak neutral currents of electrons, that couple to the weak Z bosons, are given as
\begin{equation}
J_{\mu}^Z=-\bar\psi_L\gamma^{\mu}\psi_L+2\sin^2\theta_W\,\bar\psi\gamma^{\mu}\psi\,.
\end{equation}
At low energy the Z-boson  exchange interaction can be approximated as the effective four-Fermi interaction
\begin{equation}
 {\cal L}_{\rm eff}^{NC}=-\frac{G_F}{\sqrt{2}}\,J_{\mu}^Z{J^Z}^{\mu}\,.
\end{equation}
The Z-boson exchange interaction induces the parity-violating corrections to the Cooper-pairing gap (See Fig.~\ref{wnc})
\begin{equation}
 \delta\Delta=\Delta_R-\Delta_L=\left(1-4\sin^2\theta_W\right)\cdot\frac{G_F}{\sqrt{2}}\cdot\frac{2p_F^2}{\pi^2}\frac{\Delta}{v_F}
 \cdot\ln\left(\frac{E_F}{\Delta}\right)\,\approx 1.2\times 10^{-15}\,\Delta\,,
\end{equation}
where we have taken $\sin^2\theta_W=0.23$, $E_F=p_F^2/(2m_*)=10\,{\rm eV}$, $\Delta=10^{-2}\,{\rm eV}$\,.
\section{Andreev reflection}

When an electron of energy lower than the gap ($E<\Delta$) gets scattered at the boundary of the superconductor from the normal state, it does not penetrate into the superconductor but it gets only reflected at the boundary because of the energy barrier for electrons to enter the superconductor. However electrons can transfer twice of their charge to the superconductor by so-called Andreev reflection at the boundary~\cite{andreev}. When electrons enter a superconductor, it will disappear into vacuum, leaving behind a hole with spin parallel,  but momentum opposite to the initial electron, known as Andreev reflection~\cite{andreev}, since the quantum numbers of the condensate of Cooper pairs can flow into or out of the vacuum in superconductors. 
In an ideal case the probability of Andreev reflection is one when the energy of incoming electron is less than the gap and the interface barrier has 100\% transparency. However in a non-ideal case, 
where the transparency is not 100\%,  the probability is in general a function of the superconducting gap. The larger the gap is, the more electrons or holes  get Andreev-reflected~\cite{dolcini}. Therefore, if unpolarized electrons enter, the Andreev-reflected holes will be polarized due to the difference in the superconducting gaps of left-handed electrons and right-handed electrons. Namely, as $\Delta_L>\Delta_R$, the electron with spin parallel to the momentum, corresponding to the left-handed electron, will get more Andreev-reflected to holes with spin-polarized, parallel to the momentum, but moving opposite to the initial electron~\footnote{The spin-polarization of electrons (or holes) by the Andreev  reflection due to the weak parity-violation is quite tiny but might be measurable in an experimental setup that uses the STM technique on the spin-selective Andreev-reflection, similarly to the measurement of Majorana zero modes in topological superconductor~\cite{mzm}.}. Any impurity such as magnetic impurity in the sample might mimic the parity violating effect, but the impurity effect can easily cancel out if measured from both sides of the superconductor because the chirality of incident electrons depends on their momentum direction.

\begin{figure}
\includegraphics[scale=0.5]{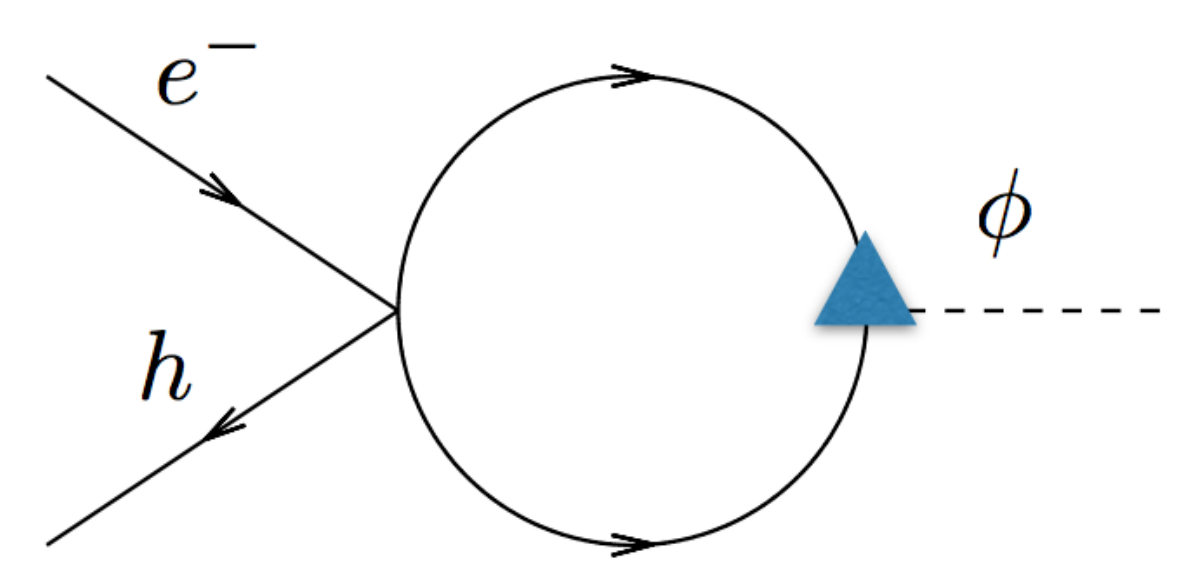}%
\caption{The  triangle denotes
the superconducting gap of electrons and $h$ denotes the hole.}
 \label{fig2}
\end{figure}

The Andreev reflection occurs at the interface of normal metals and superconducting metals.
The incoming elections, which are in the normal state, interact with each other through the four-Fermi interaction, mediated by the phonons. The phonons, emitted by the incoming electrons, will excite at the interface the gapped electrons and holes in the superconductor, which then combine to form a Cooper pair, emitting the supercurrent field $\phi(x)$. The Andreev Reflection is therefore microscopically described by the loop processes, shown in Fig.~\ref{fig2}. 
Let's suppose the interface is at the $x=0$ plane. Then, the Majorana mass term in (\ref{mass}) can be expanded as
\begin{eqnarray}
\theta(x)\Delta e^{i\phi/F}\bar\psi\psi^C&=&\theta(x)\Delta \bar\psi\psi^C+\theta(x)\,\Delta\,\frac{i\phi}{F}\bar\psi\psi^C+\cdots\,,
\end{eqnarray}
where $\phi(x)$ is the supercurrent field and $\theta (x) $ is the step function that vanishes when $x<0$ but becomes $1$ otherwise. 
Using the equation of motion for the gapped electrons, we find the coupling between the supercurrent field and the gapped electrons,
\begin{equation}
{\cal L}_{\rm y}=\frac{i}{F}\theta(x)\phi(x)\,\partial_{\mu}\left[\sum_{{\vec v}_F}\bar\psi(\vec v_F,x)\gamma^{\mu}_{\shortparallel}\psi(\vec v_F,x)\right]+\cdots\,,
\label{coupling}
\end{equation} 
where the ellipsis denotes the higher order terms in $\phi$ and $\gamma^{\mu}_{\shortparallel}=(\gamma^0,\hat v_F\hat v_F\cdot\vec\gamma)$. This new coupling in (\ref{coupling}), operator-producted with the phonon induced four-Fermi interaction in (\ref{eff}), will generate the effective Andreev vertex
\begin{equation}
{\cal L}_{ar}=i\kappa\,\delta^{\mu}(x)\,\partial_{\mu}\phi\,\bar\psi^C\sigma^2\psi\,,
\end{equation}
where $\delta^{\mu}(x)=\partial^{\mu}\theta(x)$ and the Andreev coupling at one-loop, treating the Andreev scattering perturbatively, is given as (See Fig.~\ref{fig2}.)
\begin{equation}
\kappa=\frac{g}{2\pi^2}\cdot\frac{\Delta}{F}\cdot\frac{p_F^2}{\Lambda^2}\,\ln\left(\frac{E_F}{\Delta}\right)\,.
\end{equation}

\section{New Physics}

\begin{figure}
\includegraphics[scale=0.35]{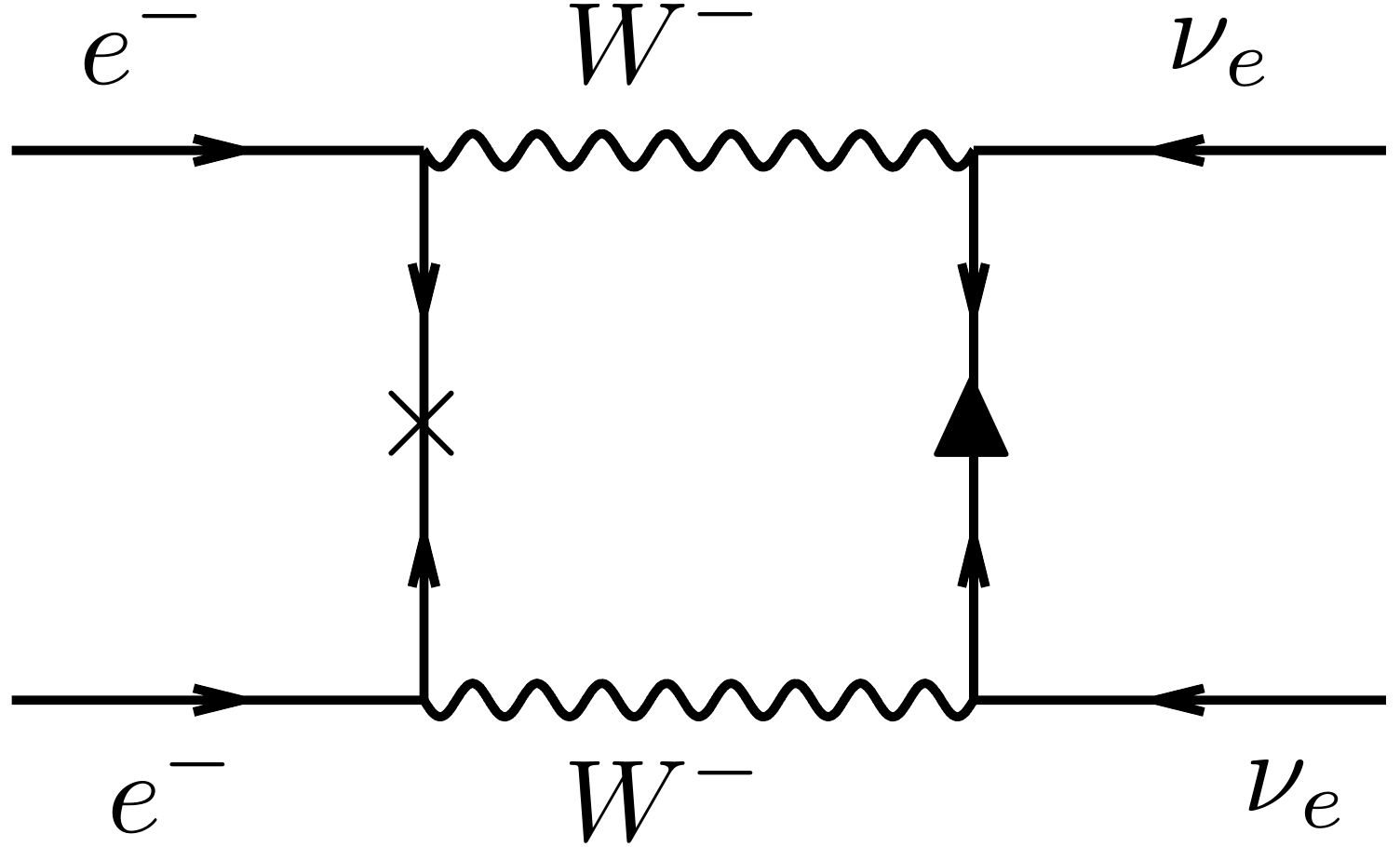}\hskip 1in
\includegraphics[scale=0.35]{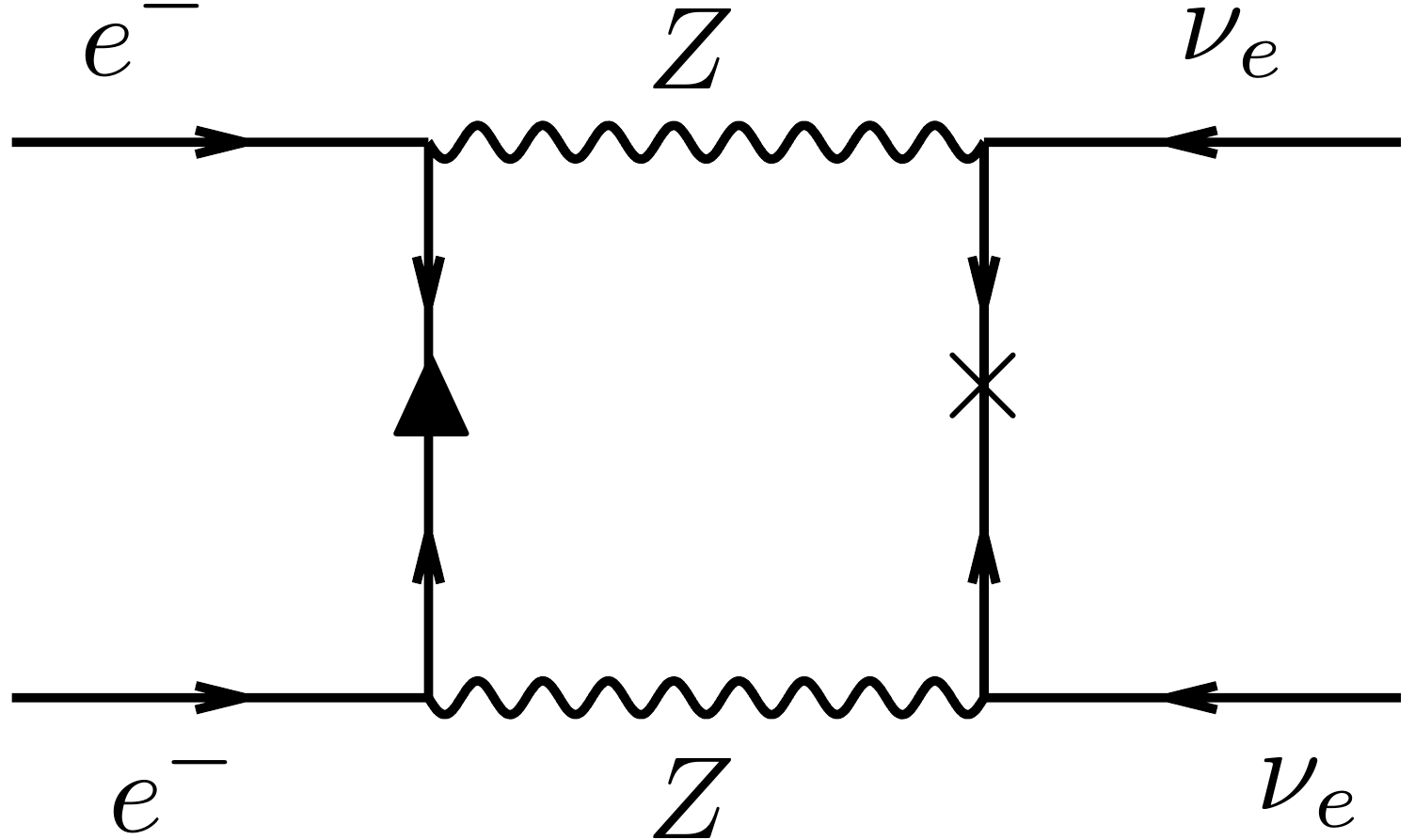}
\caption{The (one-loop) weak-interaction induced four-Fermi operators in superconductors, when active neutrinos have a Majorana mass. The  triangle denotes
the superconducting gap of electrons and the cross  denotes the Majorana mass of neutrinos. The wiggly lines denote the weak gauge bosons, $W^{\pm}$ and $Z$.}
 \label{fig3}
\end{figure}

In the effective theory of superconductors, the Majorana neutrino
mass, $m_{\nu}$, will induce by the weak interaction an operator like (see Fig.~\ref{fig3})
\begin{eqnarray}
h~ \psi_{e_L}(\vec {v}_F,x)\psi_{e_L}(-\vec
{v}_F,x){\bar\nu_{e_L}}(x){\bar\nu_{e_L}}(x)
\end{eqnarray}
where the coupling constant is found to be for $E_F\simeq 10~{\rm eV}$ 
\begin{eqnarray}
h\simeq 1.3\cdot\frac{G_F^2}{32\pi^2}\Delta\,m_{\nu}\cdot
\ln\left(\frac{E_F^2}{\Delta^2}\right)\cdot
\ln\left(\frac{E_F^2}{m_{\nu}^2}\right)
\simeq 10^{-27}\,
\frac{G_F}{\sqrt{2}}\,
\left(\frac{\Delta}{10^{-3}{\rm eV}}\right)\,
\left(\frac{m_{\nu}}{10^{-1}\,{\rm eV}}\right).
\end{eqnarray}
The interaction induced by the Majorana neutrino mass is therefore about $10^{-27}$ times weaker than the weak
interaction. This becomes the leading interaction that couples
the supercurrent to active neutrinos, if we replace the coupling
by $h\,e^{i\phi/F}$, where $\phi$ is the supercurrent field or the
Nambu-Goldstone field and $F$ is a constant analogous to the pion
decay constant.
After integrating out the gapped electrons, the effective
Lagrangian for supercurrent contains a term, given as
\begin{eqnarray}
{\cal L}_{\rm eff}\ni h\,\Delta\,p_F^2\,e^{i\phi/F}\,{\bar\nu_{e_L}}(x){\bar\nu_{e_L}}(x)
+{\rm h.c.}
\end{eqnarray}

Since the Majorana neutrino couples chirally to the electrons,
the condensate is no-longer parity-invariant. The condensate of electron Cooper-pairs
has extra contributions from the Majorana neutrino
mass, breaking parity, shown in Fig.~\ref{fig2}, 
\begin{figure}
\includegraphics[scale=0.4]{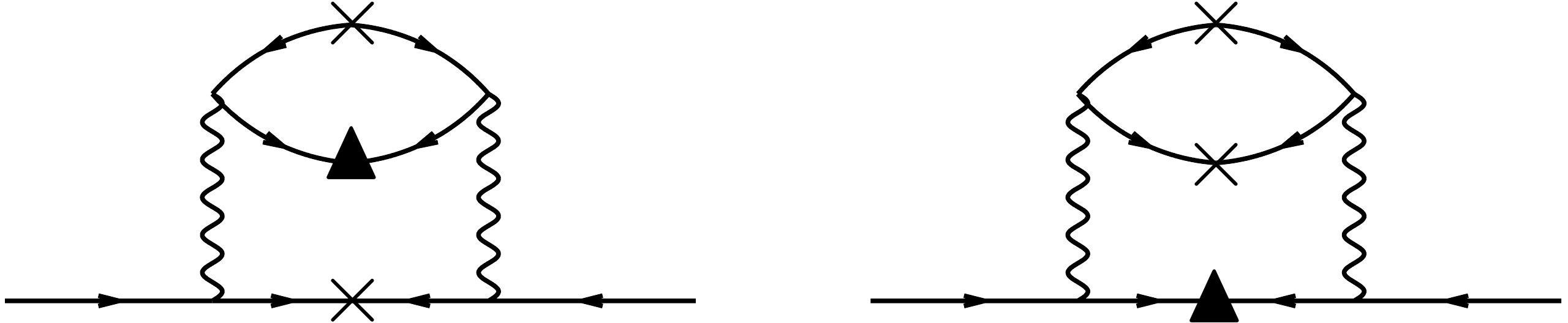}%
\caption{The extra contribution ($\delta\Delta$)
to the left-handed condensates of electrons, obtained by closing the external neutrino lines in Fig.~\ref{fig3}. The curly lines are the weak gauge bosons and the triangle (cross) denotes the superconducting gap (Majorana neutrino mass).}
 \label{fig4}
\end{figure}
\begin{equation}
\delta \Delta=\Delta_L-\Delta_R=10^{-36}\,\Delta\,\left(\frac{m_{\nu}}{0.1~{\rm eV}}\right)^2
\,\left(\frac{E_F}{10~{\rm eV}}\right)^2,
\end{equation}
where $m_{\nu}$ is the Majorana mass of the electron neutrino
and $E_F$ is the Fermi energy of the superconductor.
Any physical phenomena, such as Josephson currents~\cite{jo} or Andreev reflection~\cite{andreev}, which probe directly the superconducting gap, will be therefore sensitive to the Majorana neutrino mass. 

Finally we mention two other possible intriguing phenomena that can be seen in the measurement of the Andreev-reflection in our proposal. First one is the CP-violation in the neutrino sector. After measuring the mixing angles of three flavor neutrinos precisely the measurement of 
the CP-violating phase of PNMS matrices in the long-based line experiment is within the reach~\cite{Abe:2019vii,Pascoli:2013hxa}. If the CP-violating phase in the neutrino sector is non-vanishing, it will appear as the phase of Majorana mass term of neutrinos and therefore the Andreev-reflection probability will be different for electrons and holes to result in CP-asymmetry. Secondly, if the spin-polarization effect of the Andreev reflection is measured, it will set a stringent bound on certain beyond standard models (BSM), which have Majorana particles or some extension in the Higgs sector, coupled to electrons. 

Suppose there exists a heavy, doubly charged Higgs, as predicted in the minimal model of  type \uppercase\expandafter{\romannumeral2\relax} seesaw~\cite{Magg:1980ut,Schechter:1980gr}.
It will then generate an effective four-electron operator in superconductors, shown in Fig.~\ref{fig3}, 
\begin{figure}
\includegraphics[scale=0.6]{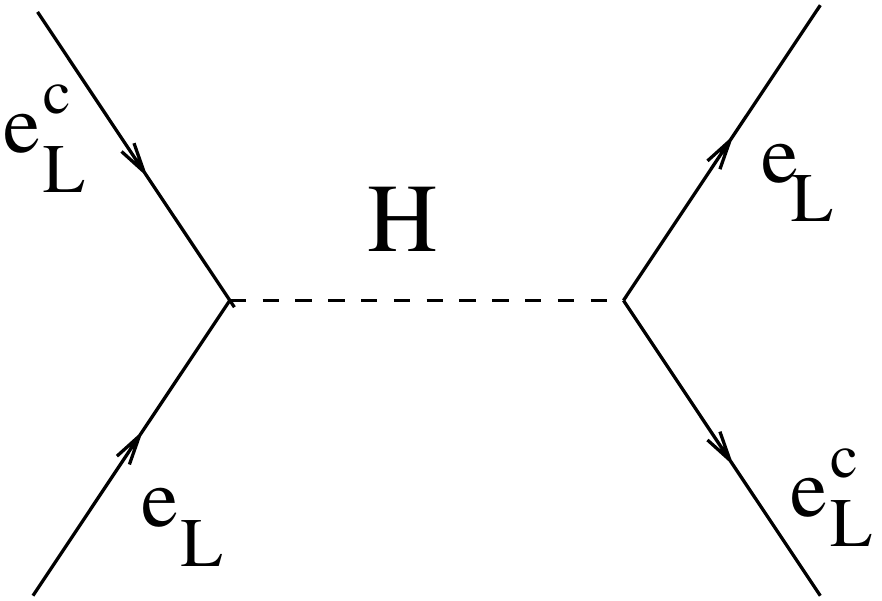}%
\caption{The four-Fermi interaction generated by the exchange of doubly charged Higgs, denoted as the dotted line.} \label{fig5}
\end{figure}
\begin{equation}
{\cal L}_{4F}=\frac{y^2}{M_H^2}\,\left(\bar{\psi_L^c}\psi_L\right)^2+{\rm h.c.}\,.
\end{equation} 
Similarly to the Majorana neutrino this effective four-electron operator due to the doubly charged Higgs will contribute to the superconducting gap of the left-handed electron (See Fig.~\ref{fig4}). Taking the Yukawa coupling $y\sim O(1)$, we get 
 \begin{equation}
\frac{\delta\Delta_L}{\Delta_L}\sim 10^{-16}\,\left(\frac{p_F}{3\,{\rm keV}}\right)^2\cdot\left(\frac{500\,{\rm GeV}}{M_H}\right)^2\,,
\end{equation} 
which turns out to be much bigger than the effect of Majorana neutrino. 
\begin{figure}
\includegraphics[scale=0.35]{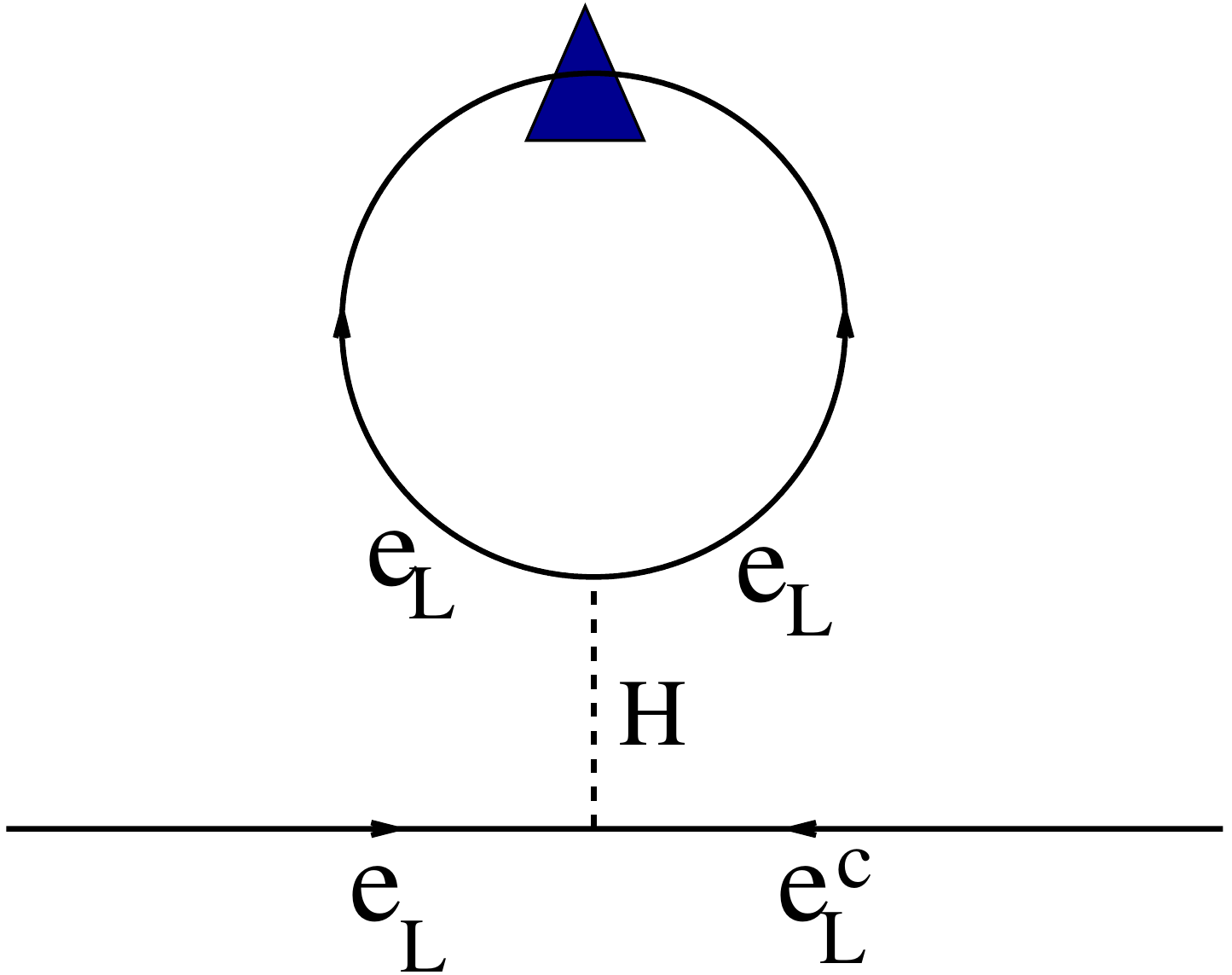}%
\caption{The superconducting gap, denoted as the triangle,  induced by the  doubly charged Higgs.} \label{fig4}
\end{figure}

To conclude we estimate the effect of the parity-violating weak interactions on the superconducting gap, which might constrain the possible new physics. 
We also find a new operator in an effective theory of superconductors, which is generated by the Majorana neutrino mass. This operator will induce additional contributions to the superconducting gap of electrons, which is nothing but the Majorana mass of electrons near the Fermi surface. Though this additional contribution to the gap is extremely small, $\delta \Delta/\Delta\sim 10^{-36}$, it might have an observable effect in physical processes that are sensitive to the superconducting gap. We find that holes or electrons are polarized by the Andreev reflection at the interface of normal metal and superconductor due to the difference in the superconducting gaps of left-handed electrons and right-handed electrons, when active neutrinos have Majorana mass. This polarization effect might be  therefore used to measure the Majorana neutrino mass.


\acknowledgments
 We thank S.~K. Kang, D.~B. Kaplan, I.~W. Kim and P. Ramond  for useful discussions.
 The author is grateful to  CERN Theory group for the hospitality during his visit, where part of this work was done. 
This research was supported by Basic Science Research Program through the National Research Foundation of Korea (NRF) funded by the Ministry of Education (NRF-2017R1D1A1B06033701).


\begin{thebibliography}{99}
\bibitem{paschos} 
  E.~A.~Paschos and L.~Wolfenstein,
  Phys.\ Rev.\ D {\bf 7}, 91 (1973).

\bibitem{Amaldi:1987fu} 
  U.~Amaldi, A.~Bohm, L.~S.~Durkin, P.~Langacker, A.~K.~Mann, W.~J.~Marciano, A.~Sirlin and H.~H.~Williams,
  Phys.\ Rev.\ D {\bf 36}, 1385 (1987).

\bibitem{Bouchiat} 
  M.~A.~Bouchiat and C.~Bouchiat,
  J.\ Phys.\ (France) {\bf 35}, 899 (1974).

\bibitem{barkov} 
  L.~M.~Barkov and M.~S.~Zolotorev,
  JETP Lett.\  {\bf 27}, 357 (1978)
  [Pisma Zh.\ Eksp.\ Teor.\ Fiz.\  {\bf 27}, 379 (1978)].
  
   
\bibitem{Noecker:1988ys} 
  M.~C.~Noecker, B.~P.~Masterson and C.~E.~Wieman,
  Phys.\ Rev.\ Lett.\  {\bf 61}, 310 (1988).

\bibitem{Wood} 
  C.~S.~Wood, S.~C.~Bennett, D.~Cho, B.~P.~Masterson, J.~L.~Roberts, C.~E.~Tanner and C.~E.~Wieman,
  Science {\bf 275}, 1759 (1997).


\bibitem{marciano_rosner} 
  W.~J.~Marciano and J.~L.~Rosner,
  Phys.\ Rev.\ Lett.\  {\bf 65}, 2963 (1990)
  [Erratum-ibid.\  {\bf 68}, 898 (1992)].
  
\bibitem{oai:arXiv.org:hep-ph/9705435} 
  A.~Deandrea,
  Phys.\ Lett.\ B {\bf 409}, 277 (1997)
  [hep-ph/9705435].
\bibitem{iw} 
  M.~I.~Gresham, I.~-W.~Kim, S.~Tulin and K.~M.~Zurek,
  Phys.\ Rev.\ D {\bf 86}, 034029 (2012)
  [arXiv:1203.1320 [hep-ph]].


\bibitem{Blundell:1990ji} 
  S.~A.~Blundell, W.~R.~Johnson and J.~Sapirstein,
  Phys.\ Rev.\ Lett.\  {\bf 65}, 1411 (1990).

\bibitem{Magg:1980ut} 
  M.~Magg and C.~Wetterich,
  Phys.\ Lett.\ B {\bf 94}, 61 (1980).

\bibitem{Schechter:1980gr} 
  J.~Schechter and J.~W.~F.~Valle,
  Phys.\ Rev.\ D {\bf 22}, 2227 (1980).

\bibitem{Hong:2000tn}
D.~K.~Hong,
Phys.\ Lett.\ B {\bf 473}, 118 (2000) [hep-ph/9812510];
%
Nucl.\ Phys.\ B {\bf 582}, 451 (2000) [hep-ph/9905523].

\bibitem{Hong:2000ng} 
  D.~K.~Hong,
  Phys.\ Rev.\ D {\bf 62}, 091501 (2000)
  [hep-ph/0006105].

\bibitem{Hong:2000ck} 
  D.~K.~Hong,
  Acta Phys.\ Polon.\ B {\bf 32}, 1253 (2001)
  [hep-ph/0101025].
  
  



\bibitem{andreev}
A. F. Andreev, Zh. Eksp. Teor. Fiz. 46, 1823 (1964)
[Sov. Phys. -J. Exp. Theor. Phys. 19, 1228 (1964)].


\bibitem{dolcini}
See for instance F. Dolcini, ``Andreev Reflection'', Lecture notes for XXIII Physics GradDays, Heidelberg, 5-9 October 2009.

\bibitem{mzm}
H.-H. Sun {\it et al.} Phys.\ Rev. \ Lett. {\bf 116}, 257003 (2016).

\bibitem{jo}
B.~D. Josephson, Phys.\ Lett.\ {\bf 1}, 251 (1962); See also
A. Barone and G. Paterno, {\it Physics and Applications of the Josephson
Effect} (John Willey \& Sons, Inc. 1982).


\bibitem{Abe:2019vii}
K.~Abe \textit{et al.} [T2K],
Nature \textbf{580}, no.7803, 339-344 (2020)
doi:10.1038/s41586-020-2177-0
[arXiv:1910.03887 [hep-ex]].

  
\bibitem{Pascoli:2013hxa} 
  S.~Pascoli,
  Nucl.\ Phys.\ Proc.\ Suppl.\  {\bf 237-238}, 141 (2013).
  


\end{thebibliography}
\end{document}